\documentclass[final]{svjour2}
\usepackage{graphicx}
\usepackage{rotating}
\usepackage{amssymb}
\usepackage{mathptmx}
\usepackage[numbers]{natbib}
\makeatletter
\journalname{Journal of Low Temperature Physics}

\bibpunct{}{}{,}{s}{}{,}

\begin{document}

\newcommand{\hdblarrow}{H\makebox[0.9ex][l]{$\downdownarrows$}-}
\title{Mass coupling and $Q^{-1}$ of impurity-limited normal $^3$He in a torsion pendulum}

\author{R.G. Bennett$^1$ \and N. Zhelev$^1$ \and A.D. Fefferman$^1$ \and K.Y. Fang$^2$ \and J. Pollanen$^2$ \and P. Sharma$^3$ \and  W.P. Halperin$^2$ \and J.M. Parpia$^1$}

\institute{1:Department of Physics and LASSP, Cornell University,\\ Ithaca NY, 14853, USA\\
Tel.:1-607-255-6060\\ Fax:1-607-255-6428\\
\email{rgb77@cornell.edu}
\\2: Department of Physics and Astronomy, Northwestern University,\\ Evanston, IL 60208, USA\\ 3: Dept of Physics, Royal Holloway, \\Egham, Surrey TW20 0EX, UK}

\date{04.10.2010}

\maketitle


\begin{abstract}

We present results of the $Q^{-1}$ and period shift, $\Delta P$, for $^3$He confined in a 98\% nominal open aerogel on a torsion pendulum. The aerogel is compressed uniaxially by 10\% along a direction aligned to the torsion pendulum axis and was grown within a 400 $\mu$m tall pancake (after compression) similar to an Andronikashvili geometry. The result is a high $Q$ pendulum able to resolve $Q^{-1}$ and mass coupling of the impurity-limited $^3$He over the whole temperature range. After measuring the empty cell background, we filled the cell above the critical point and observe a temperature dependent period shift, $\Delta P$, between 100 mK and 3 mK that is 2.9$\%$ of the period shift (after filling) at 100 mK. The $Q^{-1}$ due to the $^3$He decreases by an order of magnitude between 100 mK and 3 mK at a pressure of $0.14\pm0.03$ bar. We compare the observable quantities to the corresponding calculated $Q^{-1}$ and period shift for bulk $^3$He.
\end{abstract}

\section{Introduction}
Andronikashvili type resonators have been used to resolve the viscosity, superfluid density and other phenomena of the bulk superfluid phases of $^3$He for several decades\cite{parpia1,parpia2,hook}. However, for impurity limited or so-called ``dirty" $^3$He, most measurements (starting with \cite{porto}) have concentrated on the superfluid fraction, though there has been experimental\cite{golov} and theoretical work\cite{einzel1,higashitani} focussed on the $Q^{-1}$ as well. Most experiments have utilized a right circular cylinder as the aerogel's geometry, primarily in an effort to increase the signal to noise. However, a side effect is that there is a significant dissipative contribution to the empty cell by the losses in the aerogel ($Q^{-1}$). In addition there has been a recrudescence in the interest in uniaxially compressed and stretched aerogel\cite{bunkov1,bunkov2,davis,dmitriev1,dmitriev2}, following the theoretical proposal for the appearance of new phases\cite{Aoyama}, or at least a displacement of the A-B phase boundary from that in uncompressed aerogel. The present cell was designed to enable these investigations, but in this paper we concentrate on results in the normal state which are of general importance.

\section{Experimental details}
\begin{figure}
\begin{center}
\includegraphics[%
  width=0.75\linewidth,
  keepaspectratio]{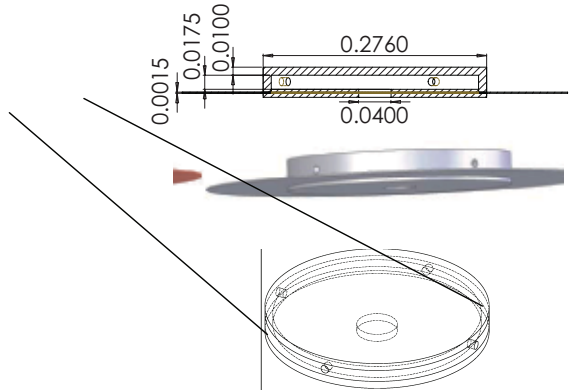}
\end{center}
\caption{On the left we show an assembly drawing of the torsional oscillator. For clarity, one of the fixed electrodes is not shown. On the right we illustrate (from top to bottom) a plan drawing of the stainless steel cavity into which the aerogel is grown (dimensions in inches) including the shim which is removed after the aerogel is dried. The middle image shows a rendition of the cavity with shim in place. The lower image shows the assembly including an opening at the bottom for thermal contact and radial holes used to optically characterize the aerogel.}
\label{Fig1}
\end{figure}

To address both of these issues (measurement of $Q^{-1}$ and the ability to resolve the shift in the P,T coordinates of the A-B phase boundary), we designed and constructed a stainless steel cavity (see Fig.~\ref{Fig1}) into which the aerogel could be grown. The cavity was fabricated with a target height (after compression of the aerogel) set at 400 $\mu$m, with an internal diameter set at 6.95 mm. The aerogel was grown into the cavity with the shims in place at Northwestern University.  After drying the shims were carefully removed and a nominal force was applied to the faces of the cavity to achieve the desired compression of 10\% along the axis of the pancake.  The aerogel was optically characterized through small holes in the cavity wall before (after) compression to verify the lack (presence) of an optical anisotropy axis as described in \cite{Pollanen1}. The cavity was then embedded in an epoxy head mounted on a double pendulum torsional oscillator. The completed assembly and illustration of the pancake cell are shown in Fig.~\ref{Fig1}.

The experiment was mounted on a PrNi$_5$ demagnetization cryostat and the temperature was monitored using a melting curve thermometer and a quartz fork resonator.\cite{fork} The fork was immersed in the same $^3$He sample which filled the torsion pendulum experiment. For the normal state experiments, the temperature scale was derived from the melting curve thermometer.

The oscillator was operated as a component of a digital phase-locked-loop in which a frequency synthesizer was used to drive the oscillator at a frequency which was held to within 300 $\mu$Hz of resonance by monitoring the out of phase component of the response using a lock-in amplifier. Non-linearities of the antisymmetric mode at 2081 Hz of the heat treated BeCu alloy torsion rod were measured at 1 mK in the empty cell response down to strains of order $10^{-9}$. Consequently, the oscillator was operated at a constant amplitude ($\pm$ 2\%) and at strains of order $10^{-9}$ in order to avoid any nonlinearities in the response over the whole temperature range. By monitoring the vector response, the drive frequency and drive amplitude we were able to continuously monitor the $Q^{-1}$ and resonant frequency of the oscillator.

The empty oscillator's resonant frequency and $Q$ were first measured between 100 mK and 1.5 mK. However, the evolving heat leak from the epoxy head meant that below about 6 mK the data flattened off since the ``virtual heat-leak" was sufficient to maintain the torsion rod above ambient temperatures. Such heat leaks have been encountered in the past\cite{Dimov1}, but are rarely accounted for in background measurements. We assumed a linear temperature dependence of the background below 6 mK. The cell was then filled at $T\ge$4 K to a pressure above the critical pressure (around 6 bar) to avoid damage to the aerogel on account of surface tension. The cell was then depressurized to $\approx$0.2 bar at 100 mK and a series of measurements were carried out to below the bulk superfluid transition. Indications of the superfluid transition were obtained by observation of a small bulk contribution ($\approx$3.0\%) of the fill signal.

\section{Results}
\begin{figure}
\begin{center}
\includegraphics[%
  width=0.87\linewidth,
  keepaspectratio]{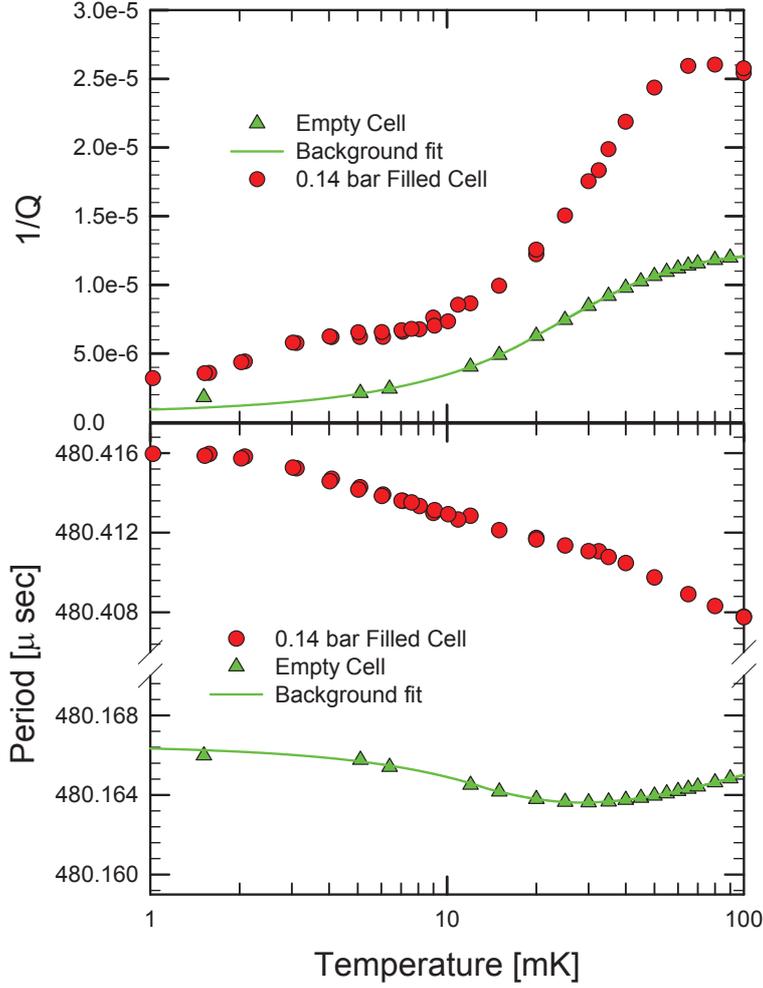}
\end{center}
\caption{(Color online) (Top Panel): The measured $Q^{-1}$ $vs$ temperature (solid triangles) for the empty oscillator, together with fit (solid line). We also plot data obtained for the cell filled with $^3$He at $0.14\pm0.03$ bar (solid circles). The fit for the empty cell data is subtracted from the filled cell data to obtain the signal from the $^3$He. (Bottom Panel): We plot the period of the empty cell (solid triangles) together with the fit through the empty cell data (solid line). The data for the cell filled with $0.14\pm0.03$ bar of $^3$He is shown as solid circles. Note the offset in the period.}
\label{Fig2}
\end{figure}
In Fig.~\ref{Fig2} we plot the measured empty cell background ($Q^{-1}$ and oscillator period). The background $Q$ ranged from $\approx$ 8$\times$10$^4$ at 100 mK to $\approx$ 5.7$\times$10$^5$ at low temperatures. The period shift showed a minimum  around 30 mK, a lower temperature than that seen in other oscillators\cite{morley}, possibly due to inadvertent variation in the heat treatments of the torsion rod. We also plot the measured $Q^{-1}$ and period for the oscillator after it was filled with $^3$He at 0.14 bar. The fits to the background are subtracted from the filled-cell data to yield the ``background subtracted data".

\begin{figure}
\begin{center}
\includegraphics[%
  width=0.95\linewidth,
  keepaspectratio]{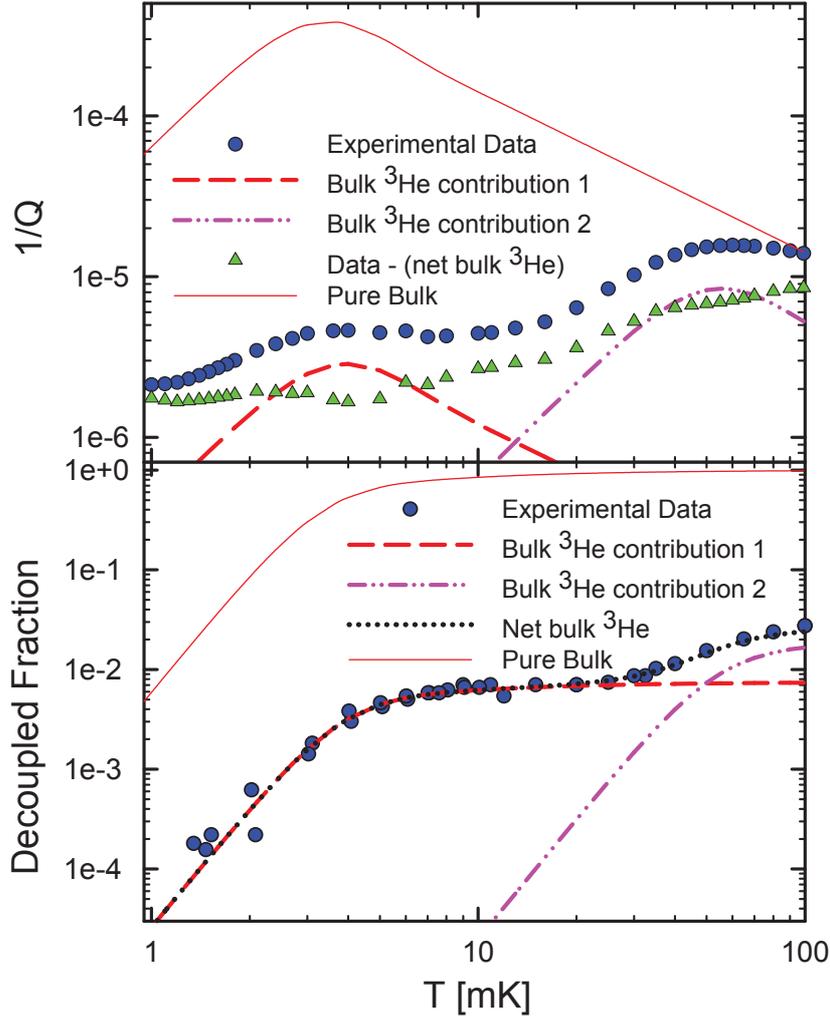}
\end{center}
\caption{(Color online) (Top Panel): The measured $Q^{-1}$ $vs$ temperature (solid data points) after the empty oscillator background is subtracted. Data was obtained at $0.14\pm0.03$ bar. For comparison, we also plot the expected $Q^{-1}$ for a 400 $\mu$m tall slab oscillator that is not filled with aerogel as the solid line. If we scale this contribution to that of a cavity of height $350\pm10$ $\mu$m but contributing only 0.75$\%$ of the inertia we obtain the dashed line (Bulk $^3$He contribution 1). A portion of the higher temperature behavior is also attributed to a bulk cavity of height $26\pm2$ $\mu$m with a contribution of 2.4$\%$ of the total inertia (Bulk $^3$He contribution 2) and is plotted as the dash-dotted line. After subtracting away the net bulk $Q^{-1}$ contributions, the $Q^{-1}$ from the $^3$He in aerogel is plotted as the solid triangles. (Bottom Panel): We plot the corresponding fraction of fluid inertia decoupled from the pendulum. The net bulk $^3$He contribution (black dots) closely follows the measured decoupled fraction. From these plots it can be seen that the inertia of $^3$He in aerogel is well-coupled to the pendulum at all temperatures, but with a small temperature independent $Q^{-1}$ and decoupled fraction above 10 mK. From the result of the fits for the inertial contributions, we can conclude that $\leq$ 3\% of the fluid inertia in the pendulum is due to bulk-like liquid (see text).}
\label{Fig3}
\end{figure}

Background subtracted data for the measured $Q^{-1}$ between 100 mK down to 1 mK are shown in Fig.~\ref{Fig3}, plotted against the temperature. Data were obtained from several temperature sweeps. It is interesting that the $Q^{-1}$ is seen to be quite small (corresponding to a $Q$ of 4$\times$10$^4$ at high temperatures before background subtraction), indicating that the $^3$He is fairly well locked to the oscillator. This observation is borne out by the period shift on filling that shows that the fraction of mass {\it decoupled} from the oscillator is very small (of order 2.9\% of the total fluid moment of inertia) at 100 mK. The fraction of inertia decoupled ($S$) is defined by the relation
\begin{equation}
    S= 1- \frac{(P(T)-P_{empty}(T))}{(P(T=0)-P_{empty}(T=0))} = 1- \frac{\Delta P(T)}{\Delta P_{max}},
\end{equation}
where $P(T)$ is the measured period at some temperature, $T, P_{empty}(T)$ is the period at the corresponding temperature with no helium added.

The results are compared to those expected for a torsion pendulum with a cavity of identical dimensions free of any impurity scatterers. The equations for the $Q^{-1}$ and inertial component have been used extensively in the past to extract the superfluid fraction and viscosity\cite{parpia3}. The calculated moment of inertia for the torsion head was 0.064 g-cm$^2$, and the fluid moment of inertia due to a cavity of dimensions 400 $\mu$m (height) $\times$ 6.95 mm (diameter) was used to obtain the estimates. We assumed that the viscosity in the Fermi liquid regime scales as $T^{-2}$, and obtained the coefficient from the results of Parpia {\it et al.}\cite{parpia4} (scaled to account for the differences in temperature scales) to generate an expression for the viscosity at this pressure $\eta = 2.05/T^{2}$[poise-mK$^2$]. We can readily see that in contrast to the expected result for the bulk fluid where virtually no fluid is coupled to the cell, the fraction of fluid inertia coupled in the aerogel-filled oscillator is close to 97\% at 100 mK. The $Q^{-1}$ of the fluid is also seen to be correspondingly reduced in the case of the aerogel filled oscillator except at 100 mK, where by coincidence it is the same as that expected for a bulk sample of 400 $\mu$m height. The $Q^{-1}$ and fluid decoupled from the pendulum for a bulk fluid sample of height 400 $\mu$m are illustrated in Fig.~\ref{Fig3} as solid lines.

We observe that the $Q^{-1}$ shows two maxima around 60 mK and 4 mK whilst the fluid decoupled from the pendulum decreases monotonically. Each of the maxima in $Q^{-1}$ appear to have characteristics that are similar to those expected for the bulk fluid, though reduced in magnitude and shifted in temperature. In order to account for these maxima and the additional fluid coupling seen below $\approx$ 6 mK, we scaled the bulk contributions to the $Q^{-1}$ and fraction decoupled, by varying the assumed height and inertia of two ``bulk cavities" located somewhere within the cell. These results are shown as the dashed and dash-dotted lines in Fig.~\ref{Fig3}. We have taken the heights of these two cavities as a free parameters to best align the $Q^{-1}$ maxima in the data to those in the bulk $Q^{-1}$ contributions. For the $Q^{-1}$ maximum at $\approx4$ mK we find that a good fit to the fluid inertia (dashed line) is obtained assuming a $350\pm10$ $\mu$m tall cavity having an inertia that is $\approx$ 0.75\% of the contribution from the $^3$He in the aerogel. Such a cavity might exist in the form of a bubble of 350 $\mu$m characteristic size, for example, in the region where the hole in the stainless steel shell meets the fill line. The higher temperature maximum in $Q^{-1}$ and decoupled fraction is best fit by a bulk cavity of height $26\pm2$ $\mu$m contributing a total of $\approx$ 2.4\% of the inertia of the $^3$He in the aerogel. Such a cavity might well exist where the face of one of the exterior stainless-steel cell walls is in contact with the epoxy head. The sum of these two bulk contributions is consistent with the $\approx$ 3\% bulk fluid signal seen when the cell, filled with $^3$He at 3 bar, is cooled below the bulk superfluid transition. No superfluid contribution from the $^3$He in aerogel is seen at 3 bar.

Models for transport in the presence of the elastic scattering channel presented by aerogel start by considering the scattering rates due to quasiparticle ($qp$) - quasiparticle scattering ($\tau^{-1}_{qp}$) that is to be summed with the elastic ({\it el}) scattering rate from the aerogel strands ($\tau^{-1}_{el}$). This latter scattering rate is the one that places the transport in the so-called ``dirty" limit, analogous to impurity scattering in normal metals. We define an effective scattering rate, $\tau^{*-1}$ = $\tau^{-1}_{qp}$ + $\tau^{-1}_{el}$. In a Fermi liquid, $\tau^{-1}_{qp}$ varies as $T^2$. Thus at low enough temperatures, the viscosity which is given by the usual relation $\eta=(1/5)nm^*v_F^2\tau_{qp}$, is modified in the dirty limit to yield $\eta=(1/5)nm^*v_F^2\tau^{*}$, since $\tau^*$ is bounded from above by the elastic scattering. Therefore at low enough temperatures, one expects the elastic scattering to dominate, and the fluid to be locked to the strands, resulting in a temperature independent $Q^{-1}$ and fluid coupled to the pendulum. A similar effect in high frequency sound attenuation was first seen and reported on by Nomura {\it et al.}.\cite{nomura} This is essentially consistent with the data shown in Fig.~\ref{Fig3}, once the bulk contributions are subtracted.

At 10 mK $\tau_{qp}$ is estimated to be 12.8 ns\cite{parpiathesis}, and if we make the assumption that $\tau_{el} < \tau_{qp}(T=10$ mK) we find that the maximum value for the scattering length in the aerogel is $l_{el}=v_F\tau_{el}<v_F\tau_{qp}=58.8\;\mathrm{[m/s]}\times12.8$ ns $=$ 776 nm, a much longer length scale than that obtained in other experiments. A more complete analysis along the lines of that presented in Venkataramani and Sauls\cite{priya} or Einzel and Parpia\cite{einzel2} would be needed to extract the velocity profile of the $^3$He in the aerogel and provide more rigorous limits for $\tau^*$ and $l_{el}$. The velocity profile is expected (at low temperatures) to match the so-called Drude limit, where the fluid velocity is not position dependent within the aerogel except for a small boundary layer at each wall. Thus, the Drude limit profile is flat across the flow channel in contrast to Hagen-Poiseuille flow which leads to a parabolic profile across the channel height. It is important to note that the treatments in the literature model the aerogel as infinitely rigid, {\it i.e.} the velocity of the strands is locked to that of the cell wall. In fact it is likely that the aerogel does move relative to the wall to some degree, generating some additional $Q^{-1}$.

Once the contribution of the bulk $^3$He is subtracted away, the remaining temperature dependent $Q^{-1}$ can only arise from the $^3$He confined in the aerogel matrix. We have adopted a simple model to account for the presence of the aerogel. Recognizing that the stiffness of the aerogel serves to lock the fluid (of reduced viscosity due to the elastic scattering mechanism) to the torsion head, we substitute for the aerogel-filled-fluid, a collection of ``slabs" equally spaced apart, that occupy the same volume as the 400 $\mu$m cell height. These slabs are filled with a Fermi fluid whose scattering rate is governed by both elastic and inelastic scattering. The slab separation and $\tau_{el}$ are considered as fitting parameters. Using this model we find that the scattering length is found to be $30\pm10$ nm, comparable to that seen in other experiments. Details of our fit and the refined model will be provided in a future publication, along with fits at a variety of pressures.

\section{Conclusions}
We have measured the $Q^{-1}$ and fraction of inertia decoupled for $^3$He within the pores of axially compressed aerogel, grown in a pancake geometry. The results show a temperature dependent fluid decoupling in the Fermi liquid region and a crossover to an impurity limited scattering rate. The characteristic scattering length is of order 30 nm, comparable to that seen in other experiments. Pressure dependent studies should allow us to better specify the scattering length and verify the magnitude of the small bulk signal.

\begin{acknowledgements}
We would like to thank E.N. Smith for technical assistance. We acknowledge support from the National Science Foundation, DMR-0806629(MWN) grant at Cornell and under DMR-0703656 at Northwestern.
\end{acknowledgements}

\pagebreak

\end{document}